\documentclass[conference]{IEEEtran}
\ifCLASSINFOpdf
  \usepackage[pdftex]{graphicx}
\else
  \usepackage[dvips]{graphicx}
\fi
\usepackage{amsmath}
\usepackage{amssymb}
\usepackage{autobreak}
\usepackage{algorithmic}

\usepackage{array}

\usepackage{stfloats}
\usepackage{url}

\usepackage{algorithm}
\usepackage{algorithmic}
\usepackage{cite}

\begin{document}

\begin{titlepage}
This work has been submitted to the IEEE for possible publication. Copyright may be transferred without notice, after which this version may no longer be accessible.

© 2020 IEEE. Personal use of this material is permitted. Permission from IEEE must be obtained for all other uses, in any current or future media, including reprinting/republishing this material for advertising or promotional purposes, creating new collective works, for resale or redistribution to servers or lists, or reuse of any copyrighted component of this work in other works.
\end{titlepage}

%
\title{Channel Estimation and Equalization for CP-OFDM-based OTFS in Fractional Doppler Channels}



\author{
\IEEEauthorblockN{Noriyuki HASHIMOTO$^\dagger$,  Noboru OSAWA$^\dagger$, Kosuke YAMAZAKI$^\dagger$, and Shinsuke IBI$^{\dagger\dagger}$\\}
\IEEEauthorblockA{$^\dagger$KDDI Research, Inc., $^{\dagger\dagger}$Doshisha University\\
\{ni-hashimoto, nb-oosawa, ko-yamazaki\}@kddi-research.jp, sibi@mail.doshisha.ac.jp}
}

\maketitle

\begin{abstract}
Orthogonal time frequency and space (OTFS) modulation is a promising technology that satisfies high-Doppler requirements for future mobile systems.
OTFS encodes information symbols and pilot symbols into the two-dimensional (2D) delay-Doppler (DD) domain. 
The received symbols suffer from inter-Doppler interference (IDI) in fading channels with fractional Doppler shifts sampled at noninteger indices in the DD domain.
The IDI has been treated as an unavoidable effect because the fractional Doppler shifts cannot be obtained directly from the received pilot symbols.
This paper provides a solution to channel estimation for fractional Doppler channels with lower computational complexity than a conventional channel estimation method using a pseudo sequence.
The proposed estimation provides new insight into the OTFS input-output relation in the DD domain as a 2D circular convolution with a small approximation.
According to the input-output relation, we also provide a low-complexity channel equalization method using the estimated channel information.
We demonstrate the error performance of the proposed channel estimation and equalization in a high-Doppler channel by simulations.
The simulation results show that the proposed channel estimation method outperforms the conventional channel estimation.
The results also show that the proposed equalization method has a similar performance to the minimum mean square error equalizer using matrix inversion.

\end{abstract}


%
\IEEEpeerreviewmaketitle

\section{Introduction}

The requirements for future mobile systems are extremely various and strict\cite{rajatheva_white_2020}. 
For example, the mobility requirement is up to 500 km/h in 5G. 
Moreover, future mobile systems will require mobility of more than 500 km/h. 
Future mobile systems are also expected to be launched for nonterrestrial coverages, e.g., underwater, unmanned-aerial-vehicles (UAVs), and low-earth-orbit (LEO) satellites. 
Due to high Doppler shifts in these environments, orthogonal frequency division multiplexing (OFDM) designed for lower-Doppler environments is no longer effective for these applications.

Orthogonal time frequency and space (OTFS) modulation is a promising technique for addressing this challenge. 
OTFS was first proposed in \cite{hadani_orthogonal_2017-1,hadani_orthogonal_2017} where the authors demonstrated that OTFS has outstanding performance compared to the performance of OFDM in high-mobility environments with higher orders of MIMO systems and millimeter-wave systems. 
Many studies have been performed on OTFS in high-mobility environments\cite{raviteja_interference_2018,kollengode_ramachandran_mimo-otfs_2018,raviteja_embedded_2019,murali_otfs_2018,shen_channel_2019} and its application to underwater acoustic channels\cite{bocus_performance_2020}. 

In OTFS, all information symbols are mapped onto the two dimensional (2D) delay-Doppler (DD) domain, and the wireless channel is also represented in the DD domain.
In \cite{raviteja_interference_2018}, unavoidable inter-Doppler interference (IDI) is reported to occur in wireless channels composed of multiple paths with the fractional Doppler that cannot be expressed in an integer index in the DD domain.
We will refer to these channels as fractional Doppler channels.
The reason why the IDI is unavoidable is that the system cannot obtain the fractional Doppler of each path directly.
The message passing (MP) algorithm is also proposed for signal detection to mitigate the IDI in the paper, but this fractional Doppler degrades the sparsity of the channel matrix and complicates the algorithm.
In principle, the IDI can be avoided if the fractional Doppler is obtained because the OTFS input-output relation in the DD domain can be expressed as a linear equation.
The general channel estimation method for OTFS is to obtain the channel response of the transmitted pilot signals in the DD domain \cite{kollengode_ramachandran_mimo-otfs_2018}.
Using this method, the system can only obtain the channel response indexed to integer numbers in the DD domain.
It is possible to mitigate the IDI effect by increasing the resolution in the Doppler domain, but this reflects on lengthening the scheduling interval of the radio resources.
In \cite{murali_otfs_2018}, the pseudo-noise (PN)-sequence-based channel estimation method is proposed to estimate the Doppler shift of each path, but the system requires a heavy computational load for accurate estimation of the fractional part.

In this paper, we propose a novel channel estimation method that can handle the fractional Doppler using the pilot response in the DD domain.
We also propose a novel low-complexity channel equalization method for OTFS by using the estimated channel information.
Our contributions can be summarized as follows:

\begin{itemize}
\item We analyze the IDI caused by the fractional Doppler and the observed channel response of pilot signals in the DD domain. Based on the analysis, we propose a novel cross-correlation-based channel estimation algorithm.
\item We formulate an OTFS input-output relation expressed as a 2D circular convolution with a small approximation derived from general channel characteristics.  We also propose a novel low-complexity channel equalization method based on this convolutional expression using the proposed channel estimation method.
\item Using simulation, we show that the OTFS system using the proposed channel estimation has better performance than the PN-sequence-based estimation and similar performance with matrix inversion type equalizer in high-mobility channels.
\end{itemize}

\textit{Notation}: Boldface capital letters represent matrices, and lower-case letters represent column vectors.  
The transpose, conjugate, conjugate transpose, and inverse of a matrix are denoted by $(\cdot)^\textrm{T}$, $(\cdot)^*$, $(\cdot)^\textrm{H}$, and $(\cdot)^{-1}$, respectively.  
$\circledast$ is the circular convolution operation. 
The operator $\textrm{vec}\{\}$ denotes the vectorization of a matrix. 
$(\cdot)_M$ is the modulo operator of divider $M$.

\section{System Model}

\begin{figure*}[t]
 \begin{center}
  \includegraphics[width=\linewidth]{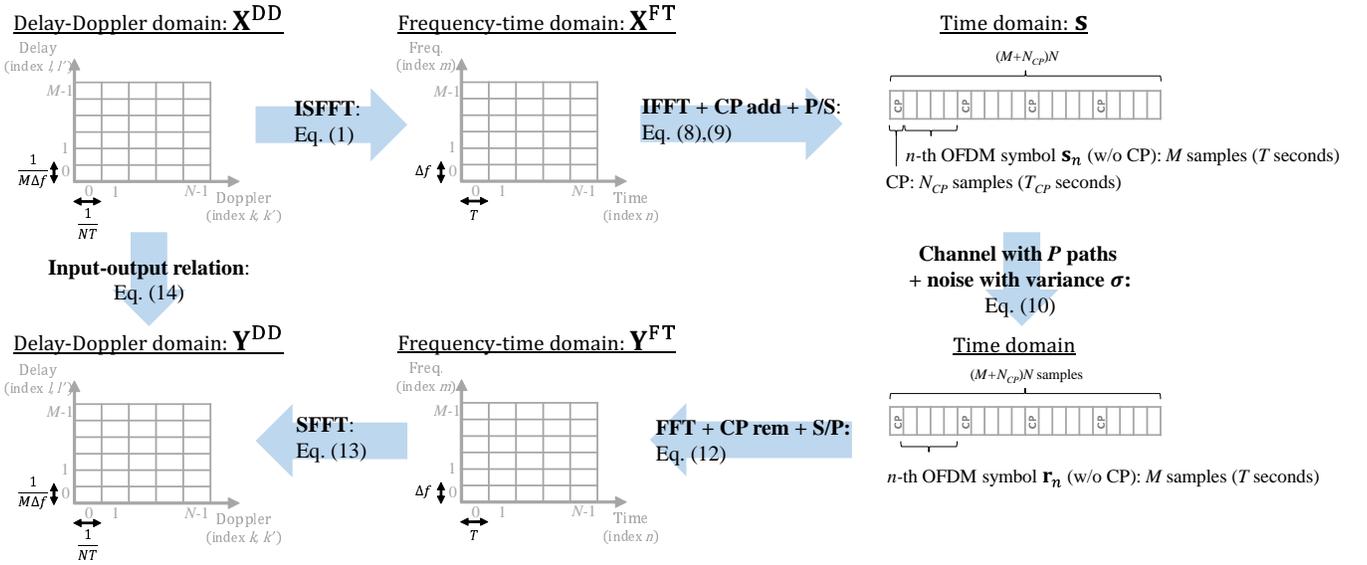}
  \caption{System model for OTFS with rectangular pulse shaping and cyclic prefix (CP-OFDM-based OTFS).}
  \label{fig:system_model}
 \end{center}
\end{figure*}

In this section, we first review the basic principle of OTFS proposed in \cite{hadani_orthogonal_2017-1}. 
Then, we explain the CP-OFDM-based OTFS system \cite{shen_channel_2019} that we consider in this paper.
The overview of the CP-OFDM-based OTFS system is shown in Fig. \ref{fig:system_model}

\subsection{Basic principle of OTFS}
In the OTFS modulation, information symbols (e.g., QAM symbols) are mapped onto the DD domain. The transmitter converts DD domain symbols $X^\textrm{DD}[l,k]$ into the frequency-time (FT) domain symbols $X^\textrm{FT}[m,n]$ using the inverse symplectic finite Fourier transform (ISFFT) as
\begin{equation}
X^\text{FT}[m,n] = \frac{1}{\sqrt{MN}}\sum_{k=0}^{N-1}\sum_{l=0}^{M-1}X^\text{DD}[l,k]e^{-j2\pi\left(\frac{ml}{M}-\frac{nk}{N}\right)}.
\end{equation}
where $M$ and $N$ are the numbers of delay and Doppler bins in the DD domain, respectively.

A time-domain transmit signal $s(t)$ is given by applying the Heisenberg transform to $X^\textrm{FT}[m,n]$ as described by
\begin{equation}
s(t) = \sum_{n=0}^{N-1}\sum_{m=0}^{M-1}X^\text{FT}[m,n] e^{j2\pi m\Delta f(t-nT)} g_\text{tx}(t-nT)
\label{eq:TransmitSignal}
\end{equation}
where $g_\textrm{tx}(t)$ is the transmit pulse shape, and $T$ and $\Delta f$ are the sampling and frequency intervals, respectively, in the FT domain.

The time-domain received signal $r(t)$ in the doubly-selective fading channels is expressed as described by
\begin{equation}
r(t) = \int\int h(\tau,\nu) e^{j2\pi\nu(t-\tau)} s(t-\tau) d\tau d\nu
\end{equation}
where $h(\tau,\nu)$ is the complex-valued Doppler-variant channel impulse response characterized by delay $\tau$ and Doppler frequency $\nu$.
Assuming a ray-based quasi-static propagation channel, $h(\tau,\nu)$ is given as
\begin{equation}
    h(\tau,\nu) = \sum_{p=1}^P h_p \delta(\tau-\tau_p) \delta(\nu-\nu_p)
\end{equation}
where $P$ is the number of propagation paths, and $\delta(\cdot)$ denotes the Dirac delta function.
Each path is represented by attenuation $h_p$, delay $\tau_p$, and Doppler frequency $\nu_p$ for the $p$-th path.
The delay and Doppler values for the $p$-th path is given as $\tau_p=\frac{l_{\tau_p}}{M\Delta f}$ and $\nu_p=\frac{k_{\nu_p}+\kappa_{\nu_p}}{NT}$, where $l_{\tau_p}$ and $k_{\nu_p}$ represent the integer indices of the delay and the Doppler bins in the DD domain, respectively, and $\kappa_{\nu_p}$ is the fractional Doppler that uses a noninteger index to represent $\nu_p$.
We consider fractional delays outside the system using the fractional delay filter theory \cite{wolf_fractional_2001}.

The receiver performs matched filtering with the receive pulse shape $g_\textrm{rx}(t)$. 
This operation is known as the cross-ambiguity function $A_{g_\textrm{rx},r}(\tau,\nu)$ and is given by
\begin{equation}
A_{g_\text{rx},r}(\tau,\nu) \triangleq \int e^{-j2\pi\nu(t-\tau)} g_\text{rx}^*(t-\tau) r(t) dt.
\label{eq:CrossAmbiguity}
\end{equation}
The FT domain received symbols $Y^\textrm{FT}[m,n]$ is obtained by sampling the cross-ambiguity function $A_{g_\textrm{rx},r}(\tau,\nu)$ according to
\begin{equation}
Y^\textrm{FT}[m,n]=A_{g_\textrm{rx},r}(\tau,\nu)|_{\tau=nT,\nu=m\Delta f}.
\end{equation}
This transformation from the 1D continuous signal $r(t)$ to the 2D symbols $Y^\textrm{FT}[m,n]$ is called the discrete Wigner transform and is the inversion of the Heisenberg transform.

Finally, the receiver performs the symplectic finite Fourier transform (SFFT) to obtain the DD domain received symbols $Y^\textrm{DD}[l,k]$ as described by
\begin{equation}
Y^\textrm{DD}[l,k] = \frac{1}{\sqrt{MN}}\sum_{n=0}^{N-1}\sum_{m=0}^{M-1}Y^\textrm{FT}[m,n]e^{-j2\pi\left(\frac{nk}{N}-\frac{ml}{M}\right)}.
\end{equation}

\subsection{Rectangular Pulse Shaping with Cyclic Prefix}

To simplify the basic principle of OTFS, let us define the system model in the matrix representations.
We consider the OFDM system using the cyclic prefix time guard interval (CP-OFDM) based OTFS system with a single transmit antenna and single receive antenna. 
The CP-OFDM-based OTFS system is characterized by the rectangular transmit pulse shape that is $1/\sqrt{T}$ for $t \in [0, T)$ and 0 at all other values, and rectangular receive pulse shape, which is $1/\sqrt{T}$ for $t \in [-T_\textrm{CP}, T)$ and 0 at all other values, where $T_\textrm{CP}$ is the length of CP.

By introducing the rectangular transmit pulse shape, (\ref{eq:TransmitSignal}) can be expressed by a symbol-by-symbol block $\mathbf{S} \in \mathbb{C}^{M\times N}$ as
\begin{equation}
\begin{aligned}
\mathbf{S} 
&= \frac{1}{\sqrt{T}} \mathbf{F}_M^\textrm{H} \mathbf{X}^\textrm{FT} \\ 
&= \frac{1}{\sqrt{T}} \mathbf{X}^\textrm{DD} \mathbf{F}_N^\textrm{H} 
\end{aligned}
\end{equation}
where $\mathbf{X}^\textrm{FT} \in \mathbb{C}^{M\times N}$ and $\mathbf{X}^\textrm{DD} \in \mathbb{C}^{M\times N}$ are matrices composed of $X^\textrm{FT}[m,n]$ and $X^\textrm{DD}[l,k]$, respectively, and $\mathbf{F}_M \in \mathbb{C}^{M\times M}$ and $\mathbf{F}_N \in \mathbb{C}^{N\times N}$ are the discrete Fourier transform (DFT) matrices. 
The CP-OFDM-based OTFS system adds $N_\textrm{CP}$-length CP for each OFDM symbol $\mathbf{s}_n \in \mathbb{C}^M$ that corresponds to $s(t)$ for $(n-1)T \leq t < nT$ in (\ref{eq:TransmitSignal}) and the column vector of $\mathbf{S}$ ($= [\mathbf{s}_1, \mathbf{s}_2, \ldots, \mathbf{s}_N]$), to avoid the inter-symbol interference (ISI) among the OFDM symbols via the CP addition matrix $\mathbf{A}_\textrm{CP} \in \mathbb{C}^{(M+N_\textrm{CP})\times M}$. 
Finally, the transmitter performs parallel to serial conversions to obtain a transmit signal with CPs $\mathbf{s} \in \mathbb{C}^{(M+N_\textrm{CP})N}$ in the time domain as described by

\begin{equation}
\begin{aligned}
\mathbf{s} &= \text{vec}\{\mathbf{A}_\textrm{CP}\mathbf{S}\} \\ 
&= \frac{1}{\sqrt{T}} \text{vec}\{ \mathbf{A}_\text{CP} \mathbf{X}^\text{DD} \mathbf{F}_{N}^\textrm{H} \}. 
\end{aligned}
\end{equation}

Due to the rectangular pulse shape and avoidance of the ISI with CPs, the $n$-th received OFDM symbol $\mathbf{r}_n \in \mathbb{C}^M$ after CP removal can be given by
\begin{equation}
\mathbf{r}_n = \mathbf{H}_n \mathbf{s}_n + \mathbf{z}_n
\end{equation}
where $\mathbf{H}_n \in \mathbb{C}^{M\times M}$ is the channel matrix of the $n$-th OFDM symbol, and $\mathbf{z}_n \in \mathbb{C}^{M}$ is the additive noise vector with a zero-mean complex normal distribution with variance $\sigma$.
Assuming a ray-based quasi-static propagation channel, $\mathbf{H}_n$ is expressed as
\begin{equation}
\mathbf{H}_n = \sum_{p=1}^{P} h_p e^{j\phi_p} \mathbf{\Delta}_{n,k_{\nu_p},l_{\tau_p}} \mathbf{\Pi}_{l_{\tau_p}}
\label{eq:ChannelMatrix}
\end{equation}

where $\phi_p \in [0,2\pi)$, $\mathbf{\Delta}_{n,k_{\nu_p},l_{\tau_p}} \in \mathbb{C}^{M\times M}$, and $\mathbf{\Pi}_{l_{\tau_p}} \in \mathbb{R}^{M\times M}$ denote the initial phase, the Doppler shift matrix, and the delay matrix, respectively, for the $p$-th path.
The Doppler shift matrix is expressed by $\mathbf{\Delta}_{n,k_{\nu_p},l_{\tau_p}} \triangleq \textrm{diag}[\omega^{(M+N_\textrm{CP})(n-1)+N_\mathrm{CP}-l_{\tau_p}}, \omega^{(M+N_\textrm{CP})(n-1)+N_\mathrm{CP}-l_{\tau_p}+1}, .., \\ \omega^{(M+N_\textrm{CP})n-l_{\tau_p}-1}]$ where $\omega=e^{\frac{j2\pi(k_{\nu_p}+\kappa_{\nu_p})}{(M+N_\mathrm{CP})N}}$.
The delay matrix $\mathbf{\Pi}_{l_{\tau_p}}$ is the forward cyclic shifted permutation matrix according to delay $l_{\tau_p}$ in the DD domain.

The receiver performs $M$-point DFT on the $n$-th OFDM symbol in the time domain to obtain the frequency domain symbol vector $\mathbf{y}_n^\textrm{FT} \in \mathbb{C}^{M\times 1}$ as
\begin{equation}
    \begin{aligned} \mathbf{y}_n^\text{FT} &= \mathbf{F}_M \mathbf{r}_n \\ 
    &= \mathbf{F}_M (\mathbf{H}_n \mathbf{s}_n + \mathbf{z}_n) \\ 
    &= \mathbf{F}_M \mathbf{H}_n \mathbf{X}^\text{DD} \mathbf{f}_n^\textrm{*} + \mathbf{F}_M\mathbf{z}_n
    \end{aligned}
\end{equation}
where $\mathbf{f}_n \in \mathbb{C}^{N}$ is the $n$-th column vector of $\mathbf{F}_N$.
Let $\mathbf{Y}^\textrm{FT} \triangleq [\mathbf{y}_1^\textrm{FT}, \mathbf{y}_2^\textrm{FT}, ..., \mathbf{y}_N^\textrm{FT}] \in \mathbb{C}^{M\times N}$ denote the 2D received symbols in the FT domain. 
Applying the SFFT to $\mathbf{Y}^\textrm{FT}$, we obtain the DD domain symbol matrix $\mathbf{Y}^\textrm{DD} \in \mathbb{C}^{M\times N}$ as 
\begin{equation}
\begin{aligned} \mathbf{Y}^\text{DD} &= \mathbf{F}_M^\textrm{H} \mathbf{Y}^\text{FT} \mathbf{F}_N \\ 
&= \sum_{n=1}^N \mathbf{F}_M^\textrm{H} \mathbf{y}_n^\text{FT} \mathbf{f}_n^\textrm{T} \\ 
&= \sum_{n=1}^N \mathbf{H}_n \mathbf{X}^\text{DD} \mathbf{f}_n^* \mathbf{f}_n^\textrm{T} + \mathbf{z}_n \mathbf{f}_n^\textrm{T}. \\ \end{aligned}
\label{eq:ydd}
\end{equation}
Based on (\ref{eq:ChannelMatrix}) and (\ref{eq:ydd}), each element $Y^\textrm{DD}[l,k]$ of $\mathbf{Y}^\textrm{DD}$ without the noise term can be expressed as
\begin{equation}
\begin{aligned} Y^\text{DD}[l,k] 
=& \sum_{n=1}^N \sum_{l'=0}^{M-1} \sum_{k'=0}^{N-1} \left[ \sum_{p=1}^P h_p \delta((l-l')_M-l_{\tau_p}) e^{j\phi_p} \right. \\ 
 & \left. \cdot e^{j2\pi (k_{\nu_p}+\kappa_{\nu_p}) \frac{(M+N_\text{CP})(n-1)+N_\text{CP}-l_{\tau_p}+l}{(M+N_\text{CP})N}} \right] \\ 
 & \cdot X^\text{DD}[l',k'] e^{-j2\pi \frac{(n-1)(k-k')}{N}} \\ 
=& \sum_{l'=0}^{M-1} \sum_{k'=0}^{N-1} X^\text{DD}[l',k'] \Lambda_l[(l-l')_M,(k-k')_N] .
\end{aligned}
\label{eq:elementOfYdd}
\end{equation}
$$
\begin{aligned}
\because \Lambda_l[(l-l')_M,(k-k')_N] \triangleq&  \sum_{p=1}^P h_p e^{j\phi_p} \delta((l-l')_M-l_{\tau_p}) \\
& \cdot \Upsilon_N((k_{\nu_p}+\kappa_{\nu_p}-(k-k') )_N) \\
& \cdot   \psi_p[l] 
\end{aligned}
$$
$$
\because \Upsilon_N(x) \triangleq \sum_{n=1}^N e^{j2\pi(n-1) \frac{x}{N}}=\frac{\sin(\pi x)}{\sin\left(\pi \frac{x}{N}\right)} e^{j\pi\frac{x(N-1)}{N}}
$$
$$
\because \psi_{p}[l] \triangleq e^{j2\pi (k_{\nu_p}+\kappa_{\nu_p}) \frac{N_\text{CP}-l_{\tau_p}+l}{(M+N_\text{CP})N}}
$$

The vectorized version of $\mathbf{Y}^\textrm{DD}$ can be represented as
\begin{equation}
    \begin{aligned}
    \mathbf{y}^\textrm{DD}
    &= \text{vec}\{\mathbf{Y}^\textrm{DD}\} \\
    &= (\mathbf{F}_N \otimes \mathbf{I}_M) \mathbf{H} (\mathbf{F}_N^\textrm{H} \otimes \mathbf{I}_M) \mathbf{x}^\textrm{DD} + (\mathbf{F}_N \otimes \mathbf{I}_M) \mathbf{z} \\
    &= \mathbf{\Phi}\mathbf{x}^\textrm{DD} + (\mathbf{F}_N \otimes \mathbf{I}_M) \mathbf{z}
    \end{aligned}
    \label{eq:vectorizedYdd}
\end{equation}
where $\mathbf{H} =  \textrm{diag}[\mathbf{H}_1, \mathbf{H}_2, \ldots, \mathbf{H}_N]$, $\mathbf{x}^\textrm{DD}=\textrm{vec}\{\mathbf{X}^\textrm{DD}\}$, $\mathbf{z} = [\mathbf{z}_1^\textrm{T}, \mathbf{z}_2^\textrm{T}, \ldots, \mathbf{z}_N^\textrm{T}]^\textrm{T}$, and $\mathbf{\Phi} = (\mathbf{F}_N \otimes \mathbf{I}_M) \mathbf{H} (\mathbf{F}_N^\textrm{H} \otimes \mathbf{I}_M)$.

\section{Fractional Doppler Analysis}

Equation (\ref{eq:elementOfYdd}) indicates that the received symbols $Y^\textrm{DD}[l,k]$ are affected by the fractional Doppler through $\Upsilon_N((k_{\nu_p}+\kappa_{\nu_p}-(k-k'))_N)$.  
Considering $\Upsilon_N(k+\kappa), k \in \mathbb{Z}, \kappa \in [0, 1)$, $\Upsilon_N(k+\kappa)$ is localized at $k$ only in the absence of fractional Doppler ($\kappa = 0$). 
By contrast, nonzero values of $0 < |\Upsilon_N(k+\kappa)| < N$ appear according to $\kappa$ at any $k$ when $\kappa \neq 0$.  
Thus, this is the source of the IDI.

Let us consider the channel response $\tilde{H}_{i,j}^\textrm{DD}[l,k]$ of a pilot signal $X_{i,j}^\textrm{DD}[l,k]$ in the DD domain
\begin{equation}
X_{i,j}^\text{DD}[l,k] = \begin{cases}
1 & \textrm{if } l=i, k=j \\
0 & \mathrm{otherwise}
\end{cases}.
\end{equation}
From (\ref{eq:elementOfYdd}), the channel response of the pilot signal can be written as
\begin{equation}
\begin{aligned}
\tilde{H}_{i,j}^\text{DD}[l,k] 
=& \sum_{p=1}^P h_p \delta((l-i)_M-l_{\tau_p}) e^{j\phi_p} \psi_p[l] \\
 & \cdot \Upsilon_N((k_{\nu_p}+\kappa_{\nu_p}-(k-j))_N). 
\end{aligned}
\label{eq:PilotResponse}
\end{equation}
The pilot signal is observed from the above equation as multiple paths with an integer Doppler shift that is different from the actual Doppler shift in the Doppler domain over the fractional Doppler channels.
Since the observed Doppler shifts are different from the actual Doppler shift, the direct use of the information from (\ref{eq:PilotResponse}) for signal detection results in detection error unless the inversion operation in (\ref{eq:elementOfYdd}) is performed.


\section{Channel Estimation}

In this section, we propose a channel estimation method using the cross-correlation of $\tilde{{H}}_{i,j}^\textrm{DD}[l,k]$ with $\Upsilon_N(k+\kappa)$ across Doppler domain elements. 
As shown in (\ref{eq:elementOfYdd}), the channel can be expressed with following parameters of each path.
The proposed method estimates them.

\begin{table}[h]
  \begin{tabular}{c l}
    $h_p$ & Path gain\\
    $l_{\tau_p}$ & Path delay represented in the DD domain\\  
    $k_{\nu_p}+\kappa_{\nu_p}$ & Path (fractional) Doppler represented in the DD domain\\ 
    $\psi_p$ & Phase shift due to the Doppler shift \\ 
    $\phi_p$ & Initial phase
  \end{tabular}
\end{table}

The normalized cross-correlation function for all $k$ between $\tilde{{H}}_{i,j}^\textrm{DD}[l,k]$ on the $l$-th delay bin and $\Upsilon_N(k+\kappa)$ with parameter $\kappa$ corresponding to the fractional Doppler is given by

\begin{equation}
\begin{aligned}  
R_{H_l,\Upsilon}(k+\kappa) 
=& \frac{1}{N^2}\sum_{k'=0}^{N-1} \tilde{H}_{i,j}^\text{DD}[l,k'] \Upsilon_N^*((k'-(k+\kappa))_N). 
\end{aligned}
\end{equation}

For simplicity, let us consider a channel having a single path on the $l$-th delay bin and the pilot signal in the DD domain of $i=j=0$. 
When the estimated Doppler index $\hat{k}+\hat{\kappa}$ matches the actual Doppler index $k_{\nu_p}+\kappa_{\nu_p}$ of the $p$-th path, this function reaches the highest magnitude, i.e., $|R_{H_l,\Upsilon}(\hat{k}+\hat{\kappa})|=h_p$.  
Therefore, by finding the highest magnitude of the function with $\hat{k}+\hat{\kappa}$, we can completely estimate the channel parameters for the $l$-th delay bin as follows:

\begin{equation}
\left\{
\begin{aligned}
\hat{k}_{\nu_{\hat{p}}}+\hat{\kappa}_{\nu_{\hat{p}}} &= \hat{k}+\hat{\kappa}, \\
\hat{l}_{\tau_{\hat{p}}} &= l, \\
\hat{h}_{\hat{p}} &= |R_{H_l,\Upsilon}(\hat{k}_{\nu_p}+\hat{\kappa}_{\nu_p})|, \\
\hat{\psi}_{\hat{p}} &= e^{j2\pi (\hat{k}_{\nu_{\hat{p}}}+\hat{\kappa}_{\nu_{\hat{p}}}) \frac{N_\text{CP}-\hat{l}_{\tau_{\hat{p}}}}{(M+N_\text{CP})N}}, \\
e^{j\hat{\phi}_{\hat{p}}}
&= \frac{R_{H_l,\Upsilon}(\hat{k}_{\nu_{\hat{p}}}+\hat{\kappa}_{\nu_{\hat{p}}})}{|R_{H_l,\Upsilon}(\hat{k}_{\nu_{\hat{p}}}+\hat{\kappa}_{\nu_{\hat{p}}})|}\hat{\psi}_{\hat{p}}^{-1}.
\end{aligned}
\right.
\label{eq:PathEstimation}
\end{equation}

For multipath scenarios, the estimated path based on (\ref{eq:PathEstimation}) is removed from the pilot response $\tilde{H}^\textrm{DD}_{i,j}$, and then the estimation is iteratively executed as described in Algorithm \ref{alg1}.

\addtolength{\topmargin}{+0.1cm}
\begin{algorithm}[t]
\small
\caption{Proposed Channel Estimation}         
\label{alg1}                          
\begin{algorithmic}                  
\renewcommand{\algorithmicrequire}{\textbf{Input:}}
\renewcommand{\algorithmicensure}{\textbf{Output:}}
\REQUIRE Pilot response matrix $\tilde{H}_{i,j}^\textrm{DD}$, thresholds $\alpha$ and $\beta$, and noise variance $\sigma^2$
\STATE Initialize $\hat{p} \Leftarrow 1$
\FOR{each delay $l$}
\STATE Initialize a vector $\mathbf{h}' \Leftarrow \tilde{H}_{i,j}^\textrm{DD}[l,:]$
\REPEAT
\FOR{each integer Doppler $k$ and fractional Doppler $\kappa$}
\STATE Calculate $R_{h',\Upsilon}(k+\kappa)$ using $\mathbf{h}'$, and form a vector $\mathbf{R}_{h',\Upsilon}$
\ENDFOR
\STATE Find $\hat{k}+\hat{\kappa}$ with the maximum value of $|\mathbf{R}_{h',\Upsilon}|$
\IF{$|R_{h',\Upsilon}(\hat{k}+\hat{\kappa})| > \hat{h}_{\hat{p}-1}$}
    \STATE break
\ENDIF
\STATE Estimate $\hat{k}_{\nu_{\hat{p}}}+\hat{\kappa}_{\nu_{\hat{p}}}$, $\hat{l}_{\tau_{\hat{p}}}$, $\hat{h}_{\hat{p}}$, $\hat{\psi}_{\hat{p}}$, and $\hat{\phi}_{\hat{p}}$ according to (\ref{eq:PathEstimation})
\STATE Update $\mathbf{h}' \Leftarrow \mathbf{h}' -\hat{h}_{\hat{p}} e^{j\hat{\phi}_{\hat{p}}} \hat{\psi}_{\hat{p}}[\hat{l}_{\tau_{\hat{p}}}] \mathbf{\Upsilon}_N(\hat{k}_{\nu_{\hat{p}}}+\hat{\kappa}_{\nu_{\hat{p}}})$
\STATE Update ${\hat{p}} \Leftarrow {\hat{p}}+1$
\UNTIL{Stopping criteria of (\ref{eq:StoppingCriteria})}
\ENDFOR
\ENSURE Path parameters $\hat{k}_{\nu_{\hat{p}}}+\hat{\kappa}_{\nu_{\hat{p}}}$, $\hat{l}_{\tau_{\hat{p}}}$, $\hat{h}_{\hat{p}}$, and $\hat{\phi}_{\hat{p}}$ for all $\hat{p}$
\end{algorithmic}
\end{algorithm}

The estimation algorithm for delay $l$ stops when $|R_{H',\Upsilon}(\hat{k}+\hat{\kappa})|$ for the estimated Doppler index $\hat{k}+\hat{\kappa}$ satisfies at least one of the following criteria:
\begin{equation}
\left| R_{H',\Upsilon}(\hat{k}+\hat{\kappa}) \right| < \begin{cases} \alpha \left| \sum_{i=1}^N \tilde{H}_{i,j}^\textrm{DD}[l,i] \right| \\ \beta \sigma. \end{cases}
\label{eq:StoppingCriteria}
\end{equation}
where $\alpha$ and $\beta$ are tuning factors that affect the number of paths to be estimated (let us denote by $\hat{P}$), which is reflected in the performance and computational load.
By using the fast Fourier transform (FFT) algorithm for the cross-correlation, the computational complexity of the algorithm can be expressed by $O(\hat{P} D MN\log N)$, where $D$ is the resolution of the fractional Doppler $\kappa$ to find, e.g, $D=10$ (=0.1 interval).
Once the tuning factors are optimized, it does not require to re-tune them.
However, the number of paths to be estimated can be varying since it depends on the channel and noise level.
The computational load can be controlled by limiting the upper bound of the number of paths to be estimated if there is a restriction in the computational load.

\section{Channel Equalization}
\label{section:ChannelEqualization}

Based on (\ref{eq:elementOfYdd}), $\Lambda_l[(l-l')_M,(k-k')_N]$ can be regarded as an effective channel in the DD domain, but it represents different channels according to the delay index $l$ because of the Doppler shift $\psi_p[l]$.  
If $\psi_p[l]$ is constant (e.g., no Doppler shift) at any $l$, the received symbol matrix can be expressed as $\mathbf{Y}^\textrm{DD} = \mathbf{X}^\textrm{DD} \circledast \mathbf{\Lambda}_l$, where $\mathbf{\Lambda}_l \in \mathbb{C}^{M\times N}$ is the matrix composed of $\Lambda_l[l,k]$ for $l=0,...,M-1, k=0,...,N-1$.
Even if $\psi_p[l]$ is not constant, we can observe that differences in $\psi_p[l]$ between a delay bin $l$ and adjacent delay bins are negligibly small.
This is because the differences depend on the Doppler index $k_{\nu_p}+\kappa_{\nu_p}$ and the frame length $(M+N_\textrm{CP})N$ from (\ref{eq:elementOfYdd}), but the frame length is dominant (i.e., $\frac{k_{\nu_p}+\kappa_{\nu_p}}{(M+N_\textrm{CP})N} \ll 1$). 

In addition, $Y^\textrm{DD}[l,k]$ is composed of $X^\textrm{DD}[l',k']$ and $\Lambda_l[(l-l')_M,(k-k')_N]$ with an index $l'$ that can only be a close value to the index $l$ in practice because the channel impulse response $h(\tau,\nu)$ is usually localized (i.e., $l_{\tau_p} \ll M$) in the delay domain i.e., $\Lambda_l[(l-l')_M,(k-k')_N)] = 0$ if $(l-l')_M-l_{\tau_p} > 0$. 
These observations indicate that the $l$-th symbols of $\mathbf{Y}^\textrm{DD}$ can be represented by the 2D circular convolution of $\mathbf{X}^\textrm{DD}$ and $\mathbf{\Lambda}'_l$, which is a matrix generated from the constant phase shift $\psi_{p,l}$ defined by the delay index $l$ of $Y^\textrm{DD}[l,k]$ instead of $\psi_p[l]$ in (\ref{eq:elementOfYdd}), as 
\begin{equation}
Y^\text{DD}[l,k] \approx (\mathbf{X}^\text{DD} \circledast \mathbf{\Lambda}'_l)[l,k].
\end{equation}
Since $\mathbf{\Lambda}'_l$ can be derived from the estimated path parameters described in the previous section, we can obtain the estimated transmit symbols on delay bin $l$ by applying the 2D Wiener deconvolution \cite{dougherty_point_2001} as
\begin{equation}
\mathcal{F}(\hat{\mathbf{X}}_l^\text{DD})[\eta,\xi] =  \frac{\mathcal{F}(\mathbf{\Lambda}'_l)^*[\eta,\xi]}{|\mathcal{F}(\mathbf{\Lambda}'_l)[\eta,\xi]|^2 + \sigma^2} \mathcal{F}(\mathbf{Y}^\text{DD})[\eta,\xi],
\label{eq:WienerDeconvolution}
\end{equation}

\begin{equation}
\hat{X}^\text{DD}[l,k] = \mathcal{F}^{-1}(\mathcal{F}(\hat{\mathbf{X}}_l^\text{DD}))[l,k],
\label{eq:WienerDeconvolution2}
\end{equation}
where $\mathcal{F}(\cdot)$ denotes the 2D DFT,  $\xi = 0, ..., N-1$, and $\eta = 0, ..., M-1$.

The dominant computational complexity of this operation arises from the constructing $\mathbf{\Lambda}'_l$ rather than the deconvolution, which complexity is $O(M^2N \log MN)$.
The complexity is $O(\hat{P}M^2N^2)$ according to (\ref{eq:elementOfYdd}).
This is because $\mathbf{\Lambda}'_l$ has an $M\times N$ size matrix and has a different version of each delay $l$ at the most.  Also, the complexity of $\Upsilon$ as a part of $\mathbf{\Lambda}_l$ is $O(N)$.
By contrast, the equalization based on the matrix inversion of (\ref{eq:vectorizedYdd}) has $O(M^3N^3)$ complexity.

\textit{Note}: If channel estimation error exists in $\mathbf{\Lambda}_l$, the Wiener deconvolution can be empirically adjusted by two following options.
The noise variance term in (\ref{eq:WienerDeconvolution}) can be considered an adjustable empirical parameter chosen to balance sharpness against noise\cite{dougherty_point_2001}.
Alternatively, the denominator of (\ref{eq:WienerDeconvolution}) should be modified locally to avoid zero division or division by a small value when high SNR \cite{pitas_digital_2000}.

\section{Simulation Results}
This section presents the bit error performance of the OTFS system with the proposed channel estimation and equalization under different parameters and environments. 
Table \ref{tbl:simPars} provides relevant parameters for all evaluations. 

For the channel model, fractional delay filters using Farrow structures \cite{wolf_fractional_2001} are used to implement the fractional delay.
Each of the Doppler shift is given by Jakes' formula as
\begin{equation}
\nu_p = \nu_{\max} \cos \theta_p
\end{equation}
where $\nu_{\max}$ is the maximum Doppler shift determined by the user speed and angle-of-arrival $\theta_p$ that is uniformly distributed over $[-\pi,\pi)$.
In all simulations, the pilot signals are ideally distorted by the same channel with the information signal and the complex Gaussian noise of the same level as the information signal for simplicity.

We use $\alpha$ and $\beta$ in (\ref{eq:StoppingCriteria}) of 1/50 and 1/10, respectively.  These values show the smallest computational load, which is the smallest number of path to be estimated, within the best performance parameter sets in our prior verification.

To compare the proposed channel estimation method, we consider the PN-sequence-based estimation\cite{murali_otfs_2018} with the same length of the sequence as the information signal using a successive estimation algorithm similar to the proposed estimation method.
The stopping criteria for the algorithm is adjusted so that the number of paths to be estimated is closer between the PN-sequence-based and proposed to be a fair comparison.
In this case, we can represent the computational load of the PN-sequence-based estimation by $O(\hat{P} N_\mathrm{CP} \frac{\nu_\mathrm{max}}{\Delta f} DN^2(M+N_\mathrm{CP}))
$\footnote{The computational complexity of the cross-correlation with $(M+N_\mathrm{CP})N$ length sequences is $O((M+N_\mathrm{CP})N)$.
In order to estimate a path, it is required to calculate the cross-correlation and search for the highest value of cross-correlation among all possible Doppler shifts in $[-\nu_\mathrm{max}, \nu_\mathrm{max}]$ with an arbitrary resolution and delays up to $N_\mathrm{CP}$.
When the resolution for Doppler is the same with the proposed method, the complexity  is $O(N_\mathrm{CP}\frac{\nu_\mathrm{max}}{\Delta f}DN^2(M+N_\mathrm{CP}))$.
This process will be executed $\hat{P}$ times. 
}.

\begin{table}[tb]
  \centering
  \caption{Simulation parameters}
  \begin{tabular}{c|c}
    Parameter & Value \\  \hline  \hline
    Carrier frequency & 0.8 GHz \\ \hline
    \# of subcarriers ($M$) & 256 \\ \hline
    \# of OFDM symbols ($N$) & 14 \\ \hline
    Subcarrier spacing ($\Delta f$) & 15 kHz \\ \hline
    CP length ($N_\textrm{CP}$) & 17 samples (6.67\% of $M$) \\ \hline
    Modulation & 16-QAM \\ \hline
    Channel coding & Turbo (1/2 coding rate) \\ & w/ 8 iterations \\ \hline
    Channel profile & EVA \cite{3gpp.36.104} \\ \hline
    User speed & 500 km/h \\ \hline
    Resolution of fractional Doppler \\ to find ($D$) & 10 \\ \hline
  \end{tabular}
  \label{tbl:simPars}
\end{table}

\begin{figure}[t]
 \centering
 \includegraphics[keepaspectratio, scale=0.6]{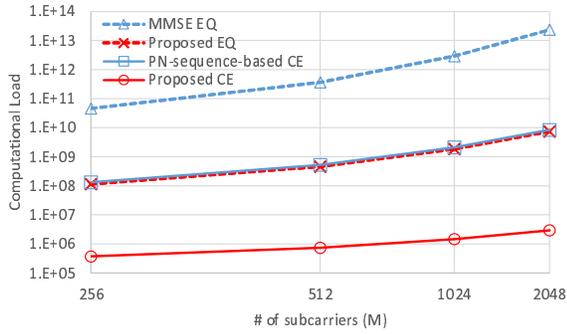}
 \caption{Comparison of computational load for the proposed channel estimation (Proposed CE), the PN-sequence-based channel estimation (PN-based CE), the proposed equalization (Proposed EQ), and the MMSE equalization using matrix inversion (MMSE EQ).}
 \label{fig:load}
\end{figure}

We first study the computational load of the proposed channel estimation and equalization compared to the PN-sequence-based channel estimation and the minimum mean square error (MMSE) equalizer based on (\ref{eq:vectorizedYdd}).
We here calculate them with the number of paths to be estimated $\hat{P}$ of 9, which is the number of paths in the EVA model.
Fig. \ref{fig:load} shows a comparison of the computational load within the parameters shown in Table \ref{tbl:simPars} with different numbers of the subcarriers.
The computational loads of the proposed channel estimation and equalization methods are much less than the conventional methods.

\begin{figure}[t]
 \centering
 \includegraphics[keepaspectratio, scale=0.65]{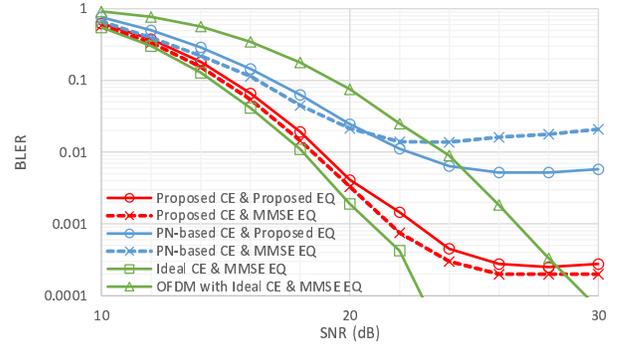}
 \caption{BLER vs. SNR (dB) for OTFS with different combinations of channel estimation and equalization methods and OFDM with the MMSE equalizer using ideal symbol-by-symbol channel impulse responses as a reference.}
 \label{fig:bler}
\end{figure}

Then, we evaluate the block error rate (BLER) performance of the proposed channel estimation and equalization methods.
Fig. \ref{fig:bler} shows that the proposed channel estimation outperforms the conventional PN-sequence-based channel estimation.
We can also see error floors in each channel estimation methods because of the channel estimation error, but the proposed estimation method is much lower.
It means that the estimation accuracy of the proposed method is much higher than the conventional one.
The proposed method is very close to the performance compared to the ideal channel estimation, except for the error floor.
Also, there is a small gap in comparing the proposed equalization with the MMSE equalization using matrix inversion.
This gap comes from the approximation in the proposed method described in section V.
This result shows that the approximation has little effect on performance.
Compared to OFDM, the result also shows that OTFS is much robust against Doppler since the equalization against Doppler shifts of each path is incorporated in operation.

\section{Conclusion}
We analyzed the effect of the fractional Doppler in the delay-Doppler domain for OTFS.
This analysis allowed us to establish a novel channel estimation method using the pilot response in the delay-Doppler domain. 
Based on the estimation, we also proposed a novel low-complexity equalizer.

Evaluation results show that the proposed estimation method displays higher channel estimation accuracy and lower computational complexity than the conventional PN-sequence-based channel estimation in a high-mobility EVA channel. 
The results also show that the proposed equalization method has lower complexity and similar performance with the MMSE equalizer using matrix inversion.





%

\bibliographystyle{ieeetr}
\bibliography{OTFS, 6G, Fundamentals}

\newpage
\section*{Appendix \\ Proof of (\ref{eq:elementOfYdd}): OTFS input-output relation in the delay-Doppler domain}

We start with defining elements of each matrix composed of (\ref{eq:elementOfYdd}).
By using the Dirac delta function, each element of the delay matrix $\mathbf{\Pi}_{l_{\tau_p}}$ can be written as
\begin{equation}
    \Pi_{l_{\tau_p}}[l,l'] = \delta((l-l')_M-l_{\tau_p}).
\end{equation}
Therefore, each element of $\mathbf{H}_n$ can be expressed as
\begin{equation}
\label{eq:elementOfHi}
    \begin{aligned}
        H_n[l,l'] =& \sum_{p=1}^P h_p e^{j\phi_p} \delta((l-l')_M - l_{\tau_p}) \\
        & \cdot e^{j2\pi(k_{\nu_p}+\kappa_{\nu_p})\frac{(M+N_\mathrm{CP})(n-1)+N_\mathrm{CP}-l_{\tau_p}+l}{(M+N_\mathrm{CP})N}}        
    \end{aligned}
\end{equation}
The signal part of (\ref{eq:ydd}) can be written as
\begin{equation}
\label{eq:omega}
    \sum_{n=1}^{N} \mathbf{H}_n \mathbf{X}^\text{DD} \mathbf{f}_n^* \mathbf{f}_n^T
    = \sum_{n=1}^{N} \mathbf{H}_n \mathbf{X}^\text{DD} \mathbf{\Omega}_n
\end{equation}

\begin{equation*}
    \because \Omega_n[k,k'] = e^{-j2\pi \frac{(n-1)(k-k')}{N}}
\end{equation*}

Based on (\ref{eq:elementOfHi}) and (\ref{eq:omega}), $Y^\mathrm{DD}[l,k]$ can be obtained as shown in (\ref{eq:yddproof}), which completes the proof.

\begin{figure*}[b]
\hrulefill
\begin{equation}
\label{eq:yddproof}
    \begin{aligned}
        Y^\text{DD}[l,k] &= \sum_{n=1}^N \sum_{l'=0}^{M-1} \sum_{k'=0}^{N-1} 
        H_n[l,l'] X^\text{DD}[l',k'] \Omega_n[k,k'] \\
        &= \sum_{n=1}^N \sum_{l'=0}^{M-1} \sum_{k'=0}^{N-1} 
        \left[ \sum_{p=1}^P h_p \delta((l-l')_M-l_{\tau_p}) e^{j\phi_p} e^{j2\pi (k_{\nu_p}+\kappa_{\nu_p}) \frac{(M+N_\text{CP})(n-1)+N_\text{CP}-l_{\tau_p}+l}{(M+N_\text{CP})N}} \right] X^\text{DD}[l',k'] e^{-j2\pi \frac{(n-1)(k-k')}{N}} \\
        &= \sum_{l'=0}^{M-1} \sum_{k'=0}^{N-1} X^\text{DD}[l',k'] \sum_{p=1}^P h_p   \delta((l-l')_M-l_{\tau_p}) e^{j\phi_p} \sum_{n=1}^N e^{j2\pi (k_{\nu_p}+\kappa_{\nu_p}) \frac{(M+N_\text{CP})(n-1)+N_\text{CP}-l_{\tau_p}+l}{(M+N_\text{CP})N}} e^{-j2\pi \frac{(n-1)(k-k')}{N}} \\
        &= \sum_{l'=0}^{M-1} \sum_{k'=0}^{N-1} X^\text{DD}[l',k'] \sum_{p=1}^P  h_p   \delta((l-l')_M-l_{\tau_p}) e^{j\phi_p} e^{j2\pi (k_{\nu_p}+\kappa_{\nu_p}) \frac{N_\text{CP}-l_{\tau_p}+l}{(M+N_\text{CP})N}} \sum_{n=1}^N e^{j2\pi (k_{\nu_p}+\kappa_{\nu_p}) \frac{n-1}{N}} e^{-j2\pi \frac{(n-1)(k-k')}{N}} \\
        &= \sum_{l'=0}^{M-1} \sum_{k'=0}^{N-1} X^\text{DD}[l',k'] \sum_{p=1}^P  h_p   \delta((l-l')_M-l_{\tau_p}) e^{j\phi_p} e^{j2\pi (k_{\nu_p}+\kappa_{\nu_p}) \frac{N_\text{CP}-l_{\tau_p}+l}{(M+N_\text{CP})N}} \sum_{n=1}^N e^{j2\pi (n-1)\frac{(k_{\nu_p}+\kappa_{\nu_p})-(k-k')}{N}}\\
        &= \sum_{l'=0}^{M-1} \sum_{k'=0}^{N-1} X^\text{DD}[l',k'] \Lambda_l[(l-l')_M,(k-k')_N]        
    \end{aligned}    
\end{equation}
\vspace*{4pt}
\end{figure*}

\end{document}